\documentclass[vecphys]{svmult}
\usepackage{makeidx}         
\usepackage{graphicx}        
\usepackage{multicol}        
\usepackage[bottom]{footmisc}
\usepackage{natbib}          
\makeindex             

\begin{document}

\title*{Wind-Blown Bubbles around Evolved Stars}
\titlerunning{Wind-Blown Bubbles} 

\author{S. Jane Arthur}
\institute{Centro de Radioastronom\'{\i}a y Astrof\'{\i}sica, UNAM,
Campus Morelia, Apartado Postal 3-72, 58090 Morelia, Michoac\'an, M\'exico
\texttt{j.arthur@astrosmo.unam.mx}.}

\maketitle

Most stars will experience episodes of substantial mass loss at some
point in their lives.  For very massive stars, mass loss dominates
their evolution, although the mass loss rates are not known exactly,
particularly once the star has left the main sequence. Direct
observations of the stellar winds of massive stars can give
information on the current mass-loss rates, while studies of the ring
nebulae and HI shells that surround many Wolf-Rayet (WR) and luminous blue variable (LBV) stars provide
information on the previous mass-loss history.  The evolution of the
most massive stars, $M > 25 M_\odot$, essentially follows the sequence O
star $\rightarrow$ LBV or red supergiant
(RSG) $\rightarrow$ WR star $\rightarrow$ supernova. For
stars of mass less than $\sim 25 M_\odot$ there is no final WR stage.
During the main sequence and WR stages, the mass loss takes the form
of highly supersonic stellar winds, which blow bubbles in the
interstellar and circumstellar medium. In this way, the mechanical luminosity of the
stellar wind is converted into kinetic energy of the swept-up ambient
material, which is important for the dynamics of the interstellar
medium. In this article, analytic and numerical models are used to
describe the hydrodynamics and energetics of wind-blown bubbles. A
brief review of observations of bubbles is given, and
the degree to which theory is supported by observations is discussed. 

\section{Classical wind-blown bubbles}
\begin{figure}
\centering
\includegraphics[width=\textwidth]{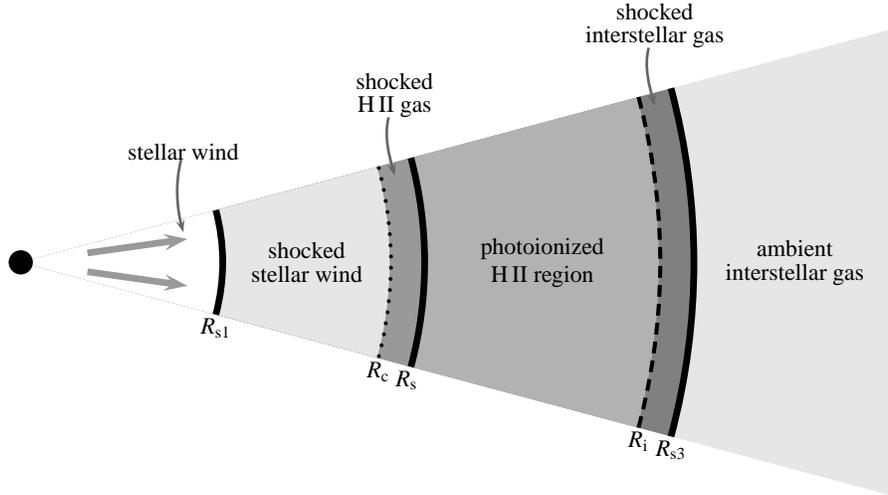}
\caption{Schematic of the different regions of a wind-blown bubble
around a massive star. $R_{\rm s1}$ is the stellar wind shock, $R_s$ is the shock set
up in the ambient medium, and $R_{\rm s3}$ is the isothermal shock sent out
ahead of the ionization front that borders the H~II region, while
$R_c$ is the contact discontinuity separating shocked stellar
wind material from swept-up ambient medium and $R_i$ marks the
ionization front, separating neutral and ionized gas.  
}
\label{fig:cartoon}
\end{figure}

The effect that a stellar wind has on its surroundings can be
dramatic. A massive star will first have formed an H~II region around
itself, and so the stellar wind interacts with this ionized gas. A
fast (typically 2000~km~s$^{-1}$) stellar wind is hypersonic with respect
to the ambient medium (sound speed $\sim$10~km~s$^{-1}$ in
photoionized gas) and so a two-shock flow pattern forms: one shock
sweeps up the ambient medium, accelerating, compressing and heating
it, while the other shock decelerates the stellar wind itself, heating
and compressing it
\citep{1968ApL.....2...97P,1972A&A....20..223D,1972SvA....15..708A}. The
basic scenario is shown in Figure~\ref{fig:cartoon}. In most
treatments, the photionized region and the neutral shell  beyond the
stellar wind bubble are not discussed separately, and it is assumed
that the shock $R_s$ expands into a fully ionized ambient medium.

Conditions behind the ambient medium
shock, $R_s$, favour rapid cooling of the swept-up shell of material, which
occurs once the shock velocity falls below 200~km~s$^{-1}$, so that
this region becomes thin and dense \citep{1975A&A....43..323F}. The shocked stellar wind, however,
can have temperature $>10^7$~K and thus does not cool efficiently. A
hot, very low density bubble of shocked stellar wind forms, which
pushes the cold, swept-up shell like a piston. Since the hot shocked
wind remains adiabatic, this is known as an ``energy-driven''
flow. 

If the ambient medium has a radial power-law density distribution $\rho(r) =
\rho_0 r^{\beta}$, and  $R_s$ and $V_s \equiv \dot{R}_s$, are the radius and velocity
of the bubble-driven shock, respectively, then the expansion law for the hot bubble can be found by considering
momentum and energy conservation, with the assumption that $R_s =
R_c$, i.e., that the region of swept-up ambient medium is thin
\citep{{1984Ap&SS.106..181D},{1989LNP...350..137D}}. The momentum and
energy equations are 
\begin{equation}
	\frac{d}{dt}\left(\frac{4\pi}{3+\beta}\rho_0 R_s^{3+\beta}V_s\right) = 4 \pi R^2 P
\label{eq:enmom}
\end{equation}
\begin{equation}
	\frac{d}{dt}\left(\frac{4\pi}{3}R_s^3\frac{P}{\gamma - 1}\right) = \dot{E}_w - 4 \pi R_s^2 P V_s
\label{eq:enene}
\end{equation}
where $P$ is the thermal pressure in the hot bubble,
giving as solution
\begin{equation}
	R_s = \phi^{1/(5 + \beta)}
\left(\frac{2\dot{E}_w}{\rho_0}\right)^{1/(5+\beta)}t^{3/(5+\beta)} \ , 
\label{eq:enrad}
\end{equation}
\begin{equation}
	V_s = \left(\frac{3}{5+\beta}\right) \frac{R_s}{t} \ , 
\label{eq:envel}
\end{equation}
with
\begin{equation}
	\phi = \frac{(3+\beta)(5+\beta)^3}{12\pi(11+\beta)(7+2\beta)}
\ .
\label{eq:enphi}
\end{equation}
Here, the stellar wind mechanical luminosity, $\dot{E}_w \equiv
\frac{1}{2}\dot{M}_wV_w$, is taken to be constant, and $\dot{M}_w$, $V_w$ are the stellar wind mass-loss rate and velocity, respectively. A time-varying
stellar wind mechanical luminosity was considered by \citet{1995ApJ...455..145G}. The case $\beta =
0$ (i.e., constant density ambient medium) returns the well-known formulae of
the thin shell approximation for stellar wind
bubble evolution \citep{1977ApJ...218..377W}. The case $\beta = -2$ would represent a wind
bubble expanding into the density distribution left by the wind of a
previous evolutionary stage of the star. 

``Momentum-driven'' flows occur when the shocked wind cools in a time
much less than the dynamical timescale. The swept-up shell of ambient
material is driven by the momentum of the wind rather than by the
pressure of the shocked wind. The expansion law can be found by
considering momentum conservation.
\begin{equation}
	\frac{d}{dt}\left(\frac{4\pi}{3}\rho_0 R_s^{3+\beta} V_s
\right) = \dot{M}_w V_w \ .
\label{eq:mommom}
\end{equation}
 For constant wind momentum, $\dot{M}_wV_w$, the radius and velocity of the outer shock are given by
\citep{{1984Ap&SS.106..181D},{1989LNP...350..137D}}
\begin{equation}
	R_s = \chi^{1/(4 + \beta)}
\left(\frac{\dot{M}_w V_w}{\rho_0}\right)^{1/(4+\beta)}t^{2/(4+\beta)} \ , 
\label{eq:momradius}
\end{equation}
\begin{equation}
	V_s = \left(\frac{2}{4+\beta}\right) \frac{R_s}{t} \ , 
\label{eq:momvel}
\end{equation}
with
\begin{equation}
	\chi = \frac{(3+\beta)(4+\beta)}{8\pi}
\ .
\label{eq:momchi}
\end{equation}

A flow can be defined as energy or momentum driven by the thermal
behaviour of the shocked stellar wind at a time, $t_0$, when the wind has
swept up its own mass of ambient material. If the shocked wind can
cool by this time then the flow will be momentum driven, otherwise it
is energy driven. The cooling time in the shocked stellar wind can be
estimated using Kahn's approximation to the cooling rate, $L = A  T^{-1/2} n^2 \ \mbox{erg~cm$^{-3}$~s$^{-1}$}$, in the
temperature range $10^5 < T < 2 \times 10^7$~K \citep{1976A&A....50..145K},
where $A = 1.3 \times 10^{-19}$ is a constant and $n$ and $T$ are the number
density and temperature, respectively, in the shocked gas. The cooling
time is thus
\begin{equation}
	t_{\rm cool} = \frac{p}{(\gamma - 1)L} \approx 4\times 10^{-35} \frac{V_w^3}{\rho_R} \ ,
\end{equation}
where a mean nucleon mass of $m_p = 2 \times 10^{-24}$~g has been used and
it is assumed that the gas is fully ionized.\footnote{The strong shock
conditions $p = \frac{3}{4} \rho_R V_w^2$, $n = 4 \rho_R/m_p$ and $T =
\frac{3}{32} V_w^2 m_p/k$ have been used in the derivation.} 
For a momentum-driven flow, $\rho_R = \dot{M}_w V_w/4\pi R_{\rm s1}^2$, where
$R_{\rm s1}$ is assumed to be the same as $R_s$ (see Eq.~\ref{eq:momradius}), since both the shocked wind and
shocked swept-up ambient medium shells are thin in this
approximation. Hence 
\begin{equation}
	t_{\rm cool} \simeq 1 \times 10^{-33} \chi^{2/(4+\beta)}
\dot{M}_w^{-(2+\beta)/(4+\beta)}V_w^{(18+4\beta)/(4+\beta)}
\rho_0^{-2/(4+\beta)} t^{2/(4+\beta)}
\end{equation}
in this case.

When the flow is energy driven, the interior pressure of the hot
bubble, $P_b$, is uniform and can be found by substituting
equations~\ref{eq:enrad} and \ref{eq:envel} into
equation~\ref{eq:enmom}. 
The density in the hot, shocked stellar wind is $n =
P_b/2kT$, where $T$ is the post-shock temperature, which depends only
on the stellar wind velocity. In this case, therefore, the cooling time is given by
\begin{equation}
	t_{\rm cool} \simeq 2 \times 10^{-35} \xi \dot{M}_w^{-(2+\beta)/(5+\beta)}V_w^{3(7+\beta)/(5+\beta)}
\rho_0^{-3/(5+\beta)} t^{(4-\beta)/(5+\beta)} \ ,
\end{equation}
where
\begin{equation}
	\xi = \frac{(5+\beta)^2(3+\beta)}{(7+2\beta)\phi^{2+\beta}} \ .
\end{equation}

The ratio of the cooling time to the dynamical time thus varies as
$t^{-\beta/(4+\beta)}$ for a momentum-driven flow and
$t^{-(1+2\beta)/(5+\beta)}$ for an energy-driven flow. While the
cooling time is greater than the dynamical time, an initially
energy-driven flow can remain energy-driven. Thus, flows can remain
energy driven if $\beta \le -\frac{1}{2}$. If $0 > \beta >
-\frac{1}{2}$, energy-driven flows can become momentum driven. An
initially momentum-driven flow can only be maintained if $\beta > 0$,
otherwise it becomes energy driven. 

Main sequence stellar winds interacting with a uniform density ambient
medium ($\beta = 0$) will normally be initially energy driven, unless
the wind itself is very dense and can cool immediately once it
shocks. The flow will remain energy driven because of the high postshock
temperatures until, eventually, this gas begins to cool. If thermal
conduction is important between the hot, shocked wind region and the
cold swept-up shell, this can lower the
temperature and raise the density in the hot bubble. This would enhance
the cooling rate and so the changeover to a momentum-driven
flow would occur earlier. 

A Wolf-Rayet star wind has a high mass-loss rate and could initially
produce a momentum-driven flow. However, the density gradient of
the red supergiant wind ($\beta = -2$) would quickly lead to a
changeover to an energy-driven flow.

Observationally, it should be possible to discriminate between
energy-driven and momentum-driven flows by determining the efficiency
factors for the conversion of stellar wind mechanical energy and
momentum into swept-up gas kinetic energy and momentum
\citep{1982ApJ...254..569T}. From the analytical model, an
energy-driven flow has a kinetic energy efficiency factor $\epsilon
= 3(5+\beta)/(11+\beta)(7+2\beta)$ and a momentum efficiency factor
$\mu = \epsilon (V_w/V_s)$, while for a momentum-driven flow the
corresponding values are $\epsilon = V_s/V_w$ and $\mu = 1$. Values of
$\epsilon \ll 0.1$ are taken to indicate a momentum-driven flow. In
practice, however, it is difficult to estimate $\epsilon$ since the
masses of both neutral and ionized components must be taken into
account \citep{1985cgd..conf..173D}. Furthermore, outer shells may no longer be driven directly
by the current stellar wind if the bubble has become depressurized.

\subsection*{Thermal conduction}
\citet{1977ApJ...218..377W} obtain a self-similar solution for the temperature
structure across the hot bubble  under the assumptions that the
pressure is constant with radius and  thermal conduction from the hot
shocked wind to the cold swept-up shell is important.
It is also assumed that radiative losses are negligible compared to
the conductive and mechanical energy fluxes. In this model,
the mass within the hot bubble is dominated by the evaporated mass
from the shell, while this mass is negligible compared with that
remaining in the shell. The self-similar solution for the temperature distribution
across the hot bubble is
\begin{equation}
 T(r) = T_b (1 - \frac{r}{R_s})^{2/5} \ , 
\end{equation}
where $T_b = a \dot{E}_w^{8/35} n_0^{2/35} t^{-6/35}$ is the
temperature in the inner part of the bubble and $a$ is a constant
which depends on the value of the conduction coefficient. This model
has been extended to include the effect of expansion in a power-law
density distribution and non-spherical (but uniform) expansion by
\citet{1995ApJ...455..145G}, but the same dependence on the similarity
variable $r/R_s$ is retained.  It has been argued
\citep[e.g.,][]{1981iuzk.book..125D} that thermal conduction will not,
in fact, be important in stellar wind bubbles since the magnetic
fields in the swept-up material will suppress it. It turns out that
the temperature and density profiles predicted by the self-similar
model of
\citet{1977ApJ...218..377W} significantly
overestimate the X-ray emission, as compared to what has been
observed, for a given set of stellar wind bubble parameters \citep{2005ApJ...633..248W}. 

\section{Numerical simulations of wind-blown bubbles}
Numerical simulations are a powerful tool for studying the
hydrodynamics of wind-blown bubbles, complementing analytical studies
since many more physical processes can be included.\footnote{Numerical
simulations solve the full gas-dynamic equations and physical
processes are generally included via source terms.} The first,
spherically symmetric, numerical studies included radiative cooling
and were able to follow the transition between the initial fully
adiabatic stage to the stage where the swept-up interstellar medium
cools radiatively \citep{1975A&A....43..323F}. Subsequent numerical
studies in two dimensions show that instabilities develop during
the evolution of wind-blown bubbles, where the nature of the
instability present depends on the evolutionary stage and physics being
considered. Kelvin-Helmholtz, Rayleigh-Taylor,
Richtmeyer-Meshkov, Vishniac, and thermal instabilities have all been
reported \citep{{1985A&A...143...59R}, {1985A&A...147..202R},
 {1985A&A...147..209R}, {1995ApJ...455..160G}, {1996A&A...305..229G}, {1996A&A...316..133G}, 
 {1995MNRAS.277...53B}, {1995MNRAS.273..443B},
 {1997MNRAS.285..387B}, {1998MNRAS.297..747S},
{2003ApJ...594..888F}}. However, care must be taken in the
identification of these instabilities, since the resolution of the
numerical grid can play an important role, as can parameters such as
the cutoff temperature below which radiative cooling is assumed equal
to zero \citep[see][ for a more detailed
discussion]{2000Ap&SS.274..243Z}. Although the most widely used analytical
models take into account thermal conduction
\citep{{1977ApJ...218..377W},{1995ApJ...455..145G}}, very few numerical
studies include this process
\citep[notable exceptions are][]{{1998NewA....3...57Z},{2000ApJ...543L..53Z},{2000Ap&SS.274..243Z}}.
The presence of even a weak magnetic field will be enough to cause
asymmetric thermal conduction and will thus lead to an asymmetric
wind-blown bubble \citep{2000Ap&SS.274..243Z}.

In this section, the evolution of a wind-blown
bubble around a $40 M_\odot$ star from the main-sequence phase, through the red supergiant phase to
the final Wolf-Rayet phase is examined by means of numerical
simulations. The first two stages are treated as one-dimensional
(spherically symmetric) simulations, while the start of the Wolf-Rayet
phase is modelled in two dimensions (cylindrical
symmetry). Photoionization is included using the method described in
\citet{2005ApJ...627..813H}. Radiative cooling and photoionization
heating are also taken into account. Thermal conduction and magnetic fields
are not included in these models. A simple three-wind model is
considered, where the stellar wind mass-loss rate and velocity is
taken to be constant during each phase and the ambient medium has
uniform density and temperature
\citep[cf.][]{{1995ApJ...455..160G}, {2005A&A...444..837V}}. 
Although a particular simulation is described here, the general
features are common to all simulations of the evolution of wind-blown
bubbles.

\subsection{Main sequence evolution}
\begin{figure}
\centering
\includegraphics[width=\textwidth]{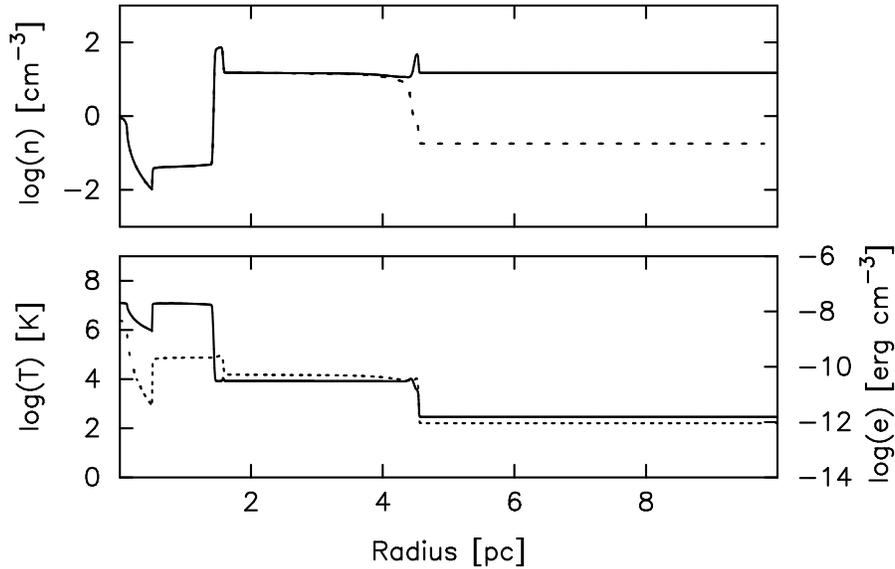}
\caption{Radial structure of a stellar wind bubble and H~II region around a $40
M_\odot $ star at an early stage of evolution (after 40,000~years).
Top panel: total
number density (solid line) and ionized number density (dotted
line). Bottom panel: temperature (solid line) and internal energy ($e
= p/[\gamma - 1]$ --- dotted line). Stellar parameters $\dot{M}_w =
9.1 \times 10^{-7} M_\odot$~yr~$^{-1}$, $V_w = 890$~km~s$^{-1}$, and $S_* = 6.34
\times 10^{47}$~s$^{-1}$ \citep[adapted from][]{2005A&A...444..837V}.}
\label{fig:early}
\end{figure}
The main-sequence evolution of the ionized region (H~II region) and stellar wind
bubble around a $40 M_\odot$ star in a uniform,
neutral ambient medium with constant stellar wind and ionizing photon
rate is presented here. The initial neutral ambient medium is taken to
have density $n_{\rm H} = 15$~cm$^{-3}$ and temperature 300~K, while
the stellar wind has a mass-loss rate of $\dot{M}_w = 9.1 \times
10^{-7} M_\odot$~yr~$^{-1}$ and wind terminal velocity $V_w =
890$~km~s$^{-1}$, and the ionizing photon rate is $S_* = 6.34
\times 10^{47}$~s$^{-1}$ \citep[cf.][]{{1992A&AS...96..269S}, {2005A&A...444..837V}}.\footnote{In these
simulations, the stellar wind is injected in the form of a thermal
energy and mass source in a small volume around the star, and for this
reason temperatures in the \textit{unshocked} stellar wind region are
high but do not affect the dynamics since the flow is highly supersonic in
this region.}

The H~II region forms immediately, reaching an initial Str\"omgren radius of
$\sim 4.5$~pc. A low-velocity, isothermal shock forms ahead of the
ionization front in the neutral gas and the hot ($T = 10^4$~K) ionized
region begins to expand. The stellar wind bubble forms inside this
ionized region. Cooling in the swept up ionized gas is very efficient
\citep[see, e.g.,][]{1975A&A....43..323F} and this region quickly forms a thin, dense
shell. The increase in opacity traps the ionizing photons and the main
ionization front becomes confined to this swept-up shell. The region
between the shell and the outer isothermal shock is occupied by gas that belonged
to the initial Str\"omgren sphere and is now slowly recombining
(recombination time is approximately $10^5/n$~years). This can be seen
in Figure~\ref{fig:early}, which shows the density, ionized density,
temperature and thermal energy (i.e., pressure) structure after
40,000~years of evolution. The outer ionization front is no longer
sharp since gas here has started to recombine. However, the shock that
was launched ahead of the ionization front continues to expand at
$< 10$~km~s$^{-1}$ into the neutral ambient medium.

In two-dimensional simulations, strong radiative cooling in the swept-up
ambient medium leads to the formation of instabilities, which cause corrugation
of the shell \citep{{1998MNRAS.297..747S}, {2003ApJ...594..888F}}.
Shadowed regions then form in the H~II region beyond the stellar wind
shell and 
recombination occurs faster here, leading to a pattern of neutral
spokes and non-radial velocity fields
\citep{{2003ApJ...594..888F}, {2005astro.ph.11035A}}. 

Eventually, the wind bubble overtakes the
outer neutral shock. By this time, the swept-up shell of material has spread out and
its density has dropped. This is because the pressure in the shell is
the same as that in the hot wind bubble, which falls slowly as the
inner wind shock moves outwards. Since the temperature in the swept-up
shell is determined by photoionization, and is thus a constant $\sim
10^4$~K, then the density has to fall (and hence the swept-up shell
must broaden) to account for the fall in pressure. The inner wind
shock will continue to move outwards until the ram pressure of the
stellar wind here ($\rho_w V_w^2 \equiv \dot{M}_w V_w/4 \pi R^2_s$)
balances the pressure of the interstellar medium. The H~II region
establishes itself in this, now, low-density shell, and sends a shock
ahead into the neutral gas. 
\begin{figure}
\centering
\includegraphics[width=\textwidth]{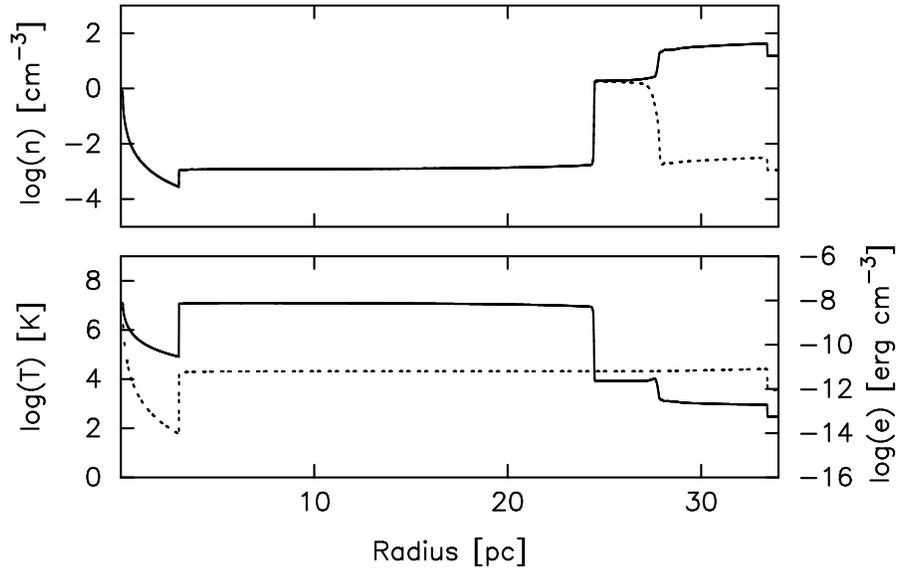}
\caption{Radial density and temperature structure of the stellar wind bubble around a $40
M_\odot $ star at the end of the main sequence stage, i.e.\ after
$\sim 4.3 \times 10^6$~years. Line types are the same as for Fig.~\ref{fig:early}.}
\label{fig:ms}
\end{figure}

Figure~\ref{fig:ms} shows the radial structure of the stellar wind bubble
and H~II region at the end of the main
sequence stage, after some $4.31 \times 10^6$~years of evolution. The
inner stellar wind shock is located about 3~pc from the star. The hot
($T \sim 10^7$~K),
very low density ($n_i \sim 10^{-3}$~cm$^{-3}$) shocked stellar wind
bubble occupies the region $3 < R < 24.5$~pc. Outside of this is the
shell containing the H~II region. The ionized gas has density $n_i
\sim 1$~cm$^{-3}$ and temperature $T = 10^4$~K, and extends between $24.5
< R < 27.5$~pc. The broad neutral shell, $27.5 < R < 33.5$~pc, has density $n_n \sim
30$~cm$^{-3}$ and temperature $T \sim 10^3$~K. The entire region
between $3 < R < 33.5$~pc has uniform pressure and is still expanding
slowly into the ambient medium.

\subsection{Post-main-sequence evolution} 
\begin{figure}
\centering
\includegraphics[width=\textwidth]{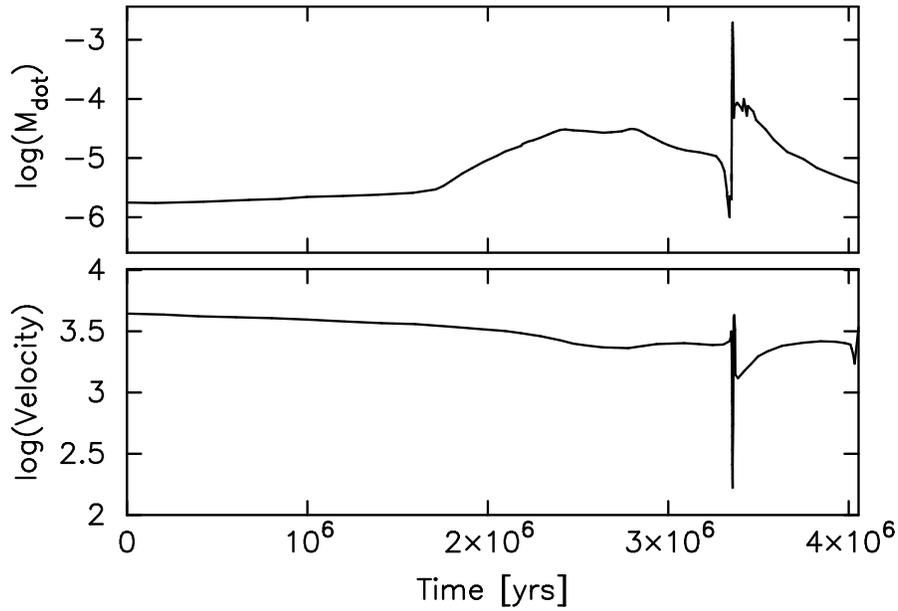}
\caption{Variation of mass-loss rate  ($M_\odot$~ yr$^{-1}$) and
stellar wind velocity (km~s$^{-1}$) as a function of time for a
$60 M_\odot$ star \citep[data from][]{1996A&A...305..229G}.}
\label{fig:lbv}
\end{figure}

In Figure~\ref{fig:lbv}, the possible mass loss history of a $60
M_\odot$ star is shown
\citep{{1994A&A...290..819L},{1996A&A...305..229G}}. A detailed study
of the evolution of the wind-blown bubble around stars of initial
masses 60~$M_\odot$ and 35~$M_\odot$ has been carried out by
\citet{{1996A&A...305..229G},{1996A&A...316..133G}}, for the purely
hydrodynamic case, and by
\citet{{2003ApJ...594..888F},{2005astro.ph.12110F}}, including the
radiative transfer of ionizing photons.  The numerical simulations
predict short-lived, observable nebulae during the LBV or RSG stage
and the onset of the WR stage. Instabilities are formed in the dense
shells swept-up by the different stellar wind stages, and these are
consistent with clumps observed in ring nebulae around some Wolf-Rayet
stars. The velocity of the LBV or RSG wind is found to play a key role
in the detailed structure of the ring nebulae formed during this
process.

\subsubsection{Red supergiant phase}
\begin{figure}
\centering
\includegraphics[width=\textwidth]{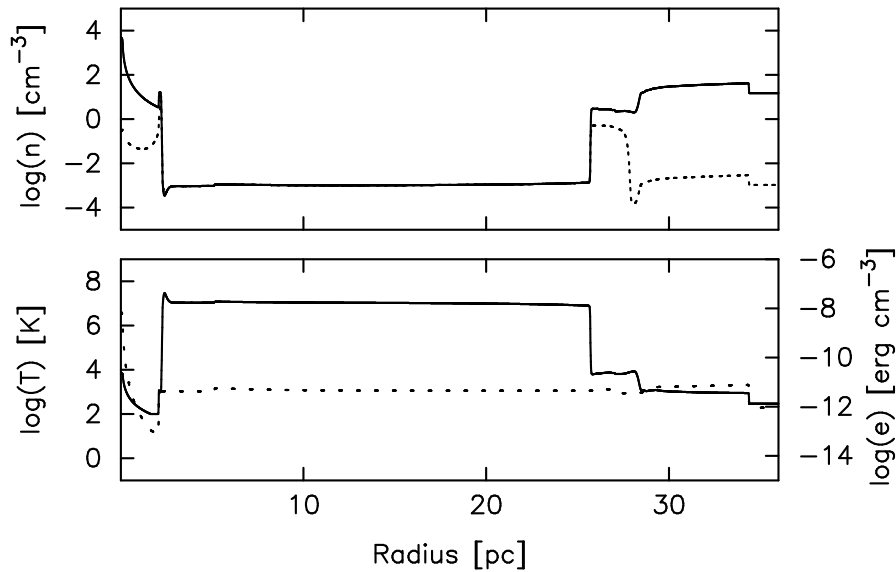}
\caption{Radial structure of a stellar wind bubble around a
40~$M_\odot$ star at the end of the red supergiant stage, i.e.\ after
$4.5\times 10^6$~years. Top panel: total
number density (solid line) and ionized number density (dotted
line). Bottom panel: temperature (solid line) and internal energy ($e
= p/[\gamma - 1]$ --- dotted line). }
\label{fig:rsg}
\end{figure}

During the red supergiant (RSG) phase, the ionizing photon luminosity
of the 40~$M_\odot$ star drops to $3.1 \times
10^{41}$~photons~s$^{-1}$, and the stellar wind velocity drops to
15~km~s$^{-1}$ while the mass-loss rate increases to $8.3 \times
10^{-5}$~$M_\odot$~yr$^{-1}$ \citep{2005A&A...444..837V}. This phase
lasts only $2 \times 10^5$~years (an LBV phase would have an even
higher mass-loss rate but a much shorter timescale, see
Fig.~\ref{fig:lbv}). The slow, dense RSG wind expands into the
structure formed by the main sequence wind and a thin, dense shell of
shocked RSG material forms ahead of the freely expanding wind (see
Figure~\ref{fig:rsg}).  The high thermal pressure in the hot main
sequence bubble causes backflow of low-density gas towards the star,
which shocks against the RSG wind. The reduction in the ionizing
photon rate leads to the disappearance of the H~II region as the shell
of ionized gas recombines. The hot bubble remains ionized since the
recombination times are long in this hot, very low-density gas. The
RSG wind itself is neutral.

\subsubsection{Wolf-Rayet phase}
\begin{figure}
\centering
\includegraphics[width=0.6\textwidth]{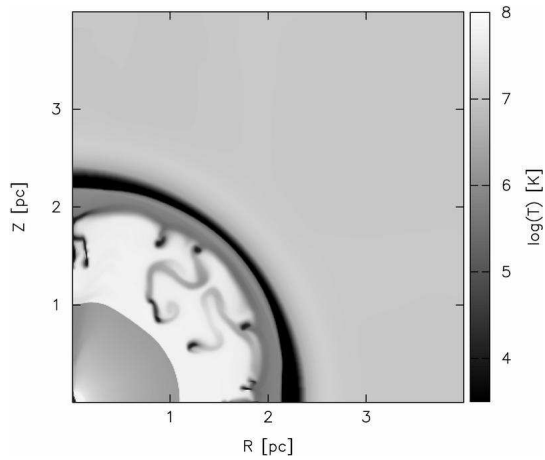}
\caption{Formation of Rayleigh-Taylor instabilities in the Wolf-Rayet
wind shell just before the interaction with the red supergiant shell. The instabilities form as the WR
wind accelerates into the $r^{-2}$ density distribution left by the RSG
wind. The temperature
of the gas is shown 8000~years after the start of the Wolf-Rayet
wind. Only one quadrant is shown, and the star is located at $(0,0)$.}
\label{fig:wr}
\end{figure}
The Wolf-Rayet wind expands into the structure formed during the
previous two phases. During the Wolf-Rayet phase, the stellar wind
becomes faster ($V_w = 2160$~km~s$^{-1}$) and remains strong
($\dot{M}_w = 4.1\times 10^{-5}$~$M_\odot$~yrs$^{-1}$) and the stellar
ionizing photon rate also increases ($S_* = 3.86 \times
10^{47}$~s$^{-1}$).  The transition between the RSG and WR winds will
involve a period of wind acceleration into the $r^{-2}$ density
distribution left by the RSG wind, and this situation is
Rayleigh-Taylor unstable.\footnote{In the numerical simulations
presented here, acceleration of the wind occurs initially as a consequence
of the way in which the stellar
wind is included in the calculation.} The WR shell consists of swept-up
RSG wind material and shocked WR wind. It becomes Rayleigh-Taylor
unstable at an early time in the simulation, well before the
interaction with the RSG shell. Dense, cold clumps of material formed
by the instability lag behind the main WR shell but the shell itself
is not disrupted, since the instabilities cease to grow once the
acceleration stage is over (see Fig.~\ref{fig:wr}).

\begin{figure}
\centering
\includegraphics[width=0.5\textwidth]{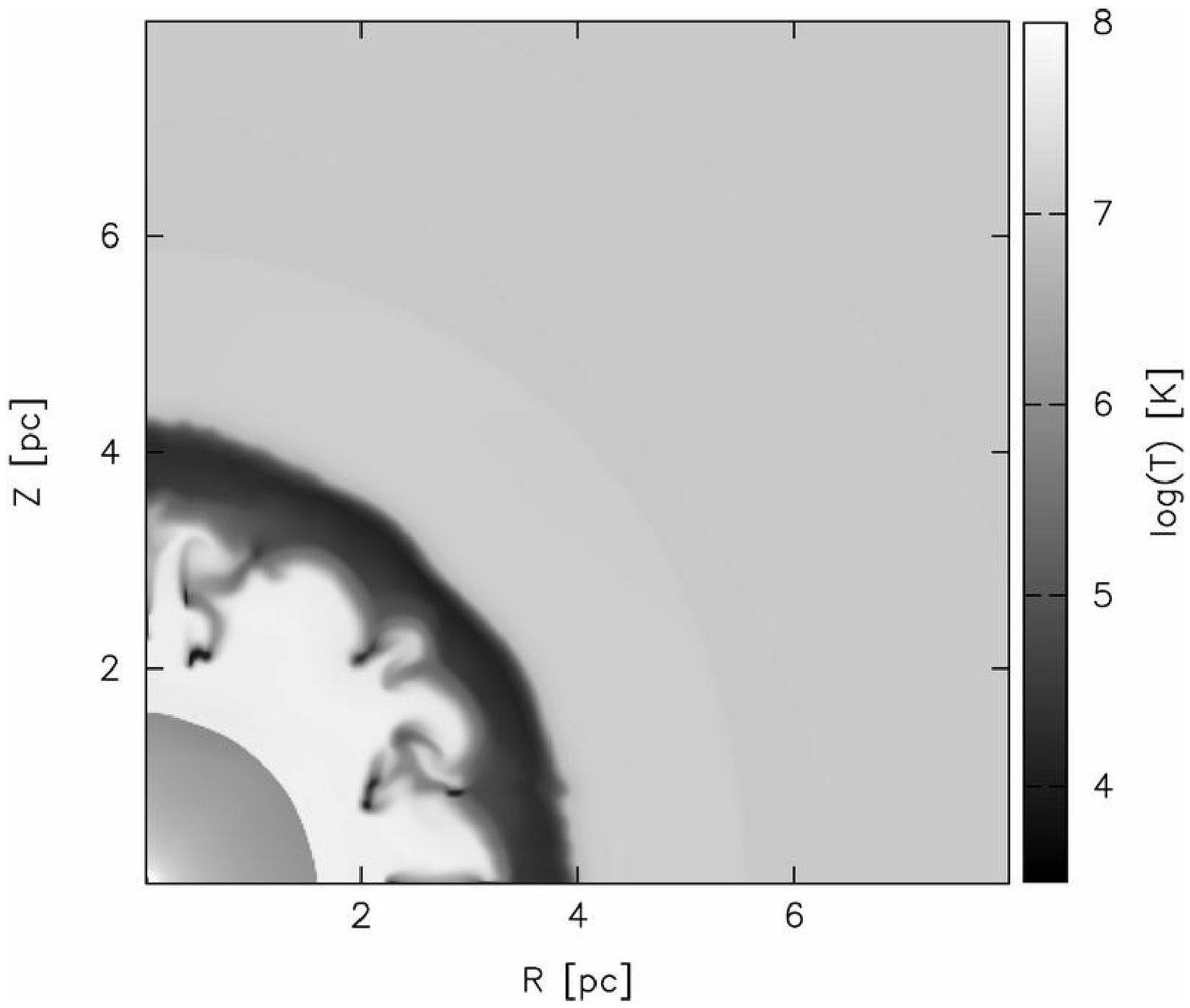}~
\includegraphics[width=0.5\textwidth]{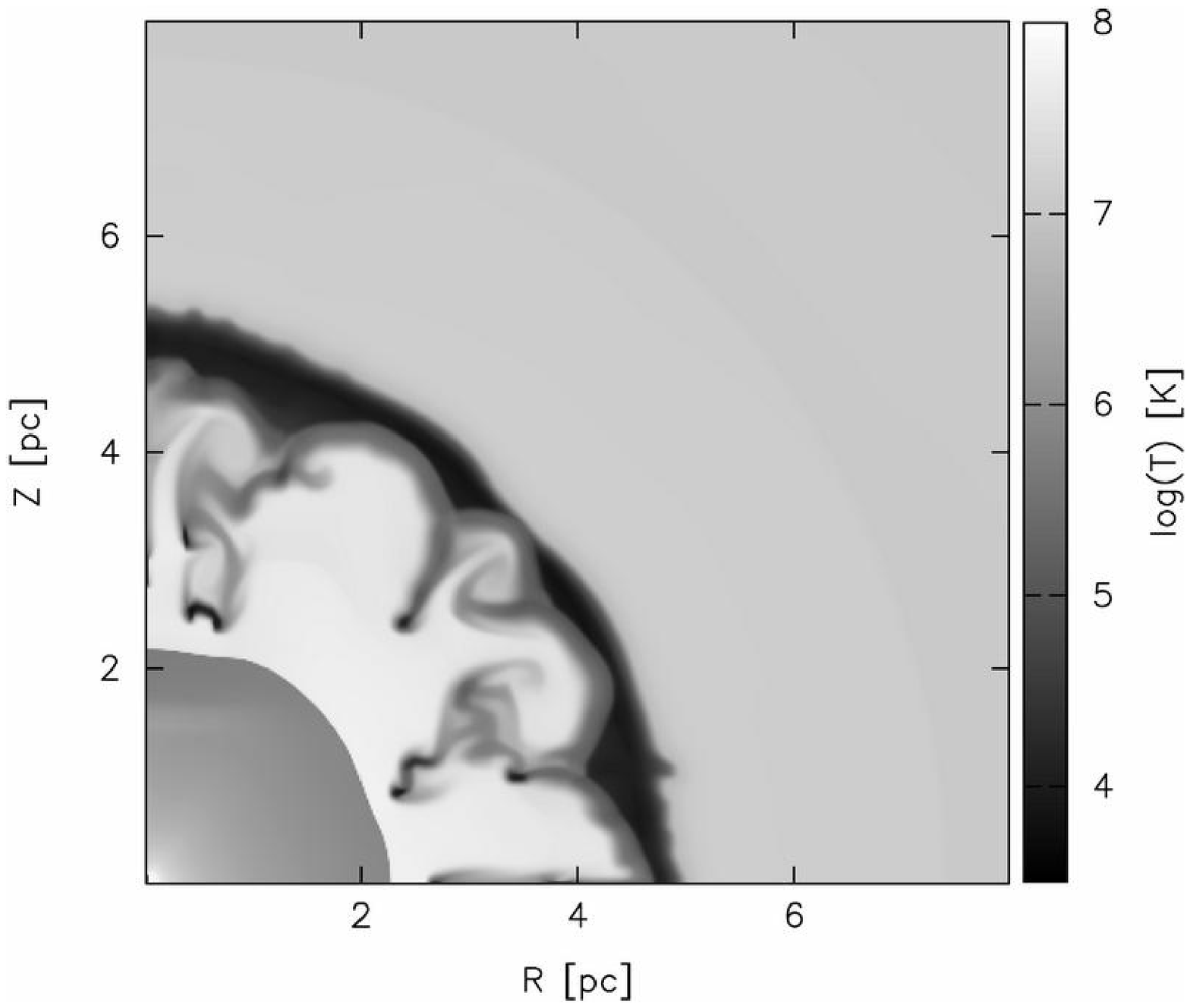}\\%
\includegraphics[width=0.5\textwidth]{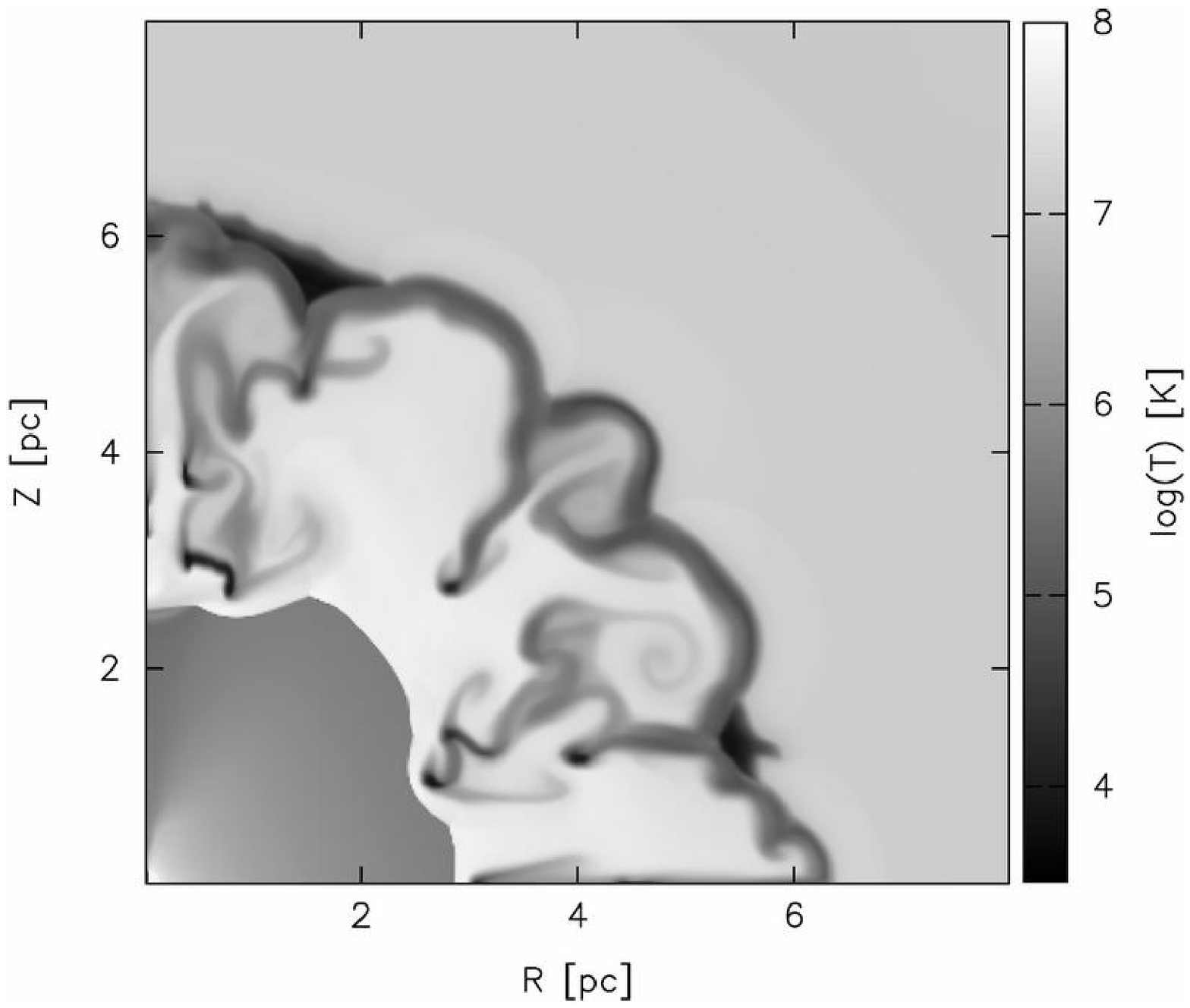}~
\includegraphics[width=0.5\textwidth]{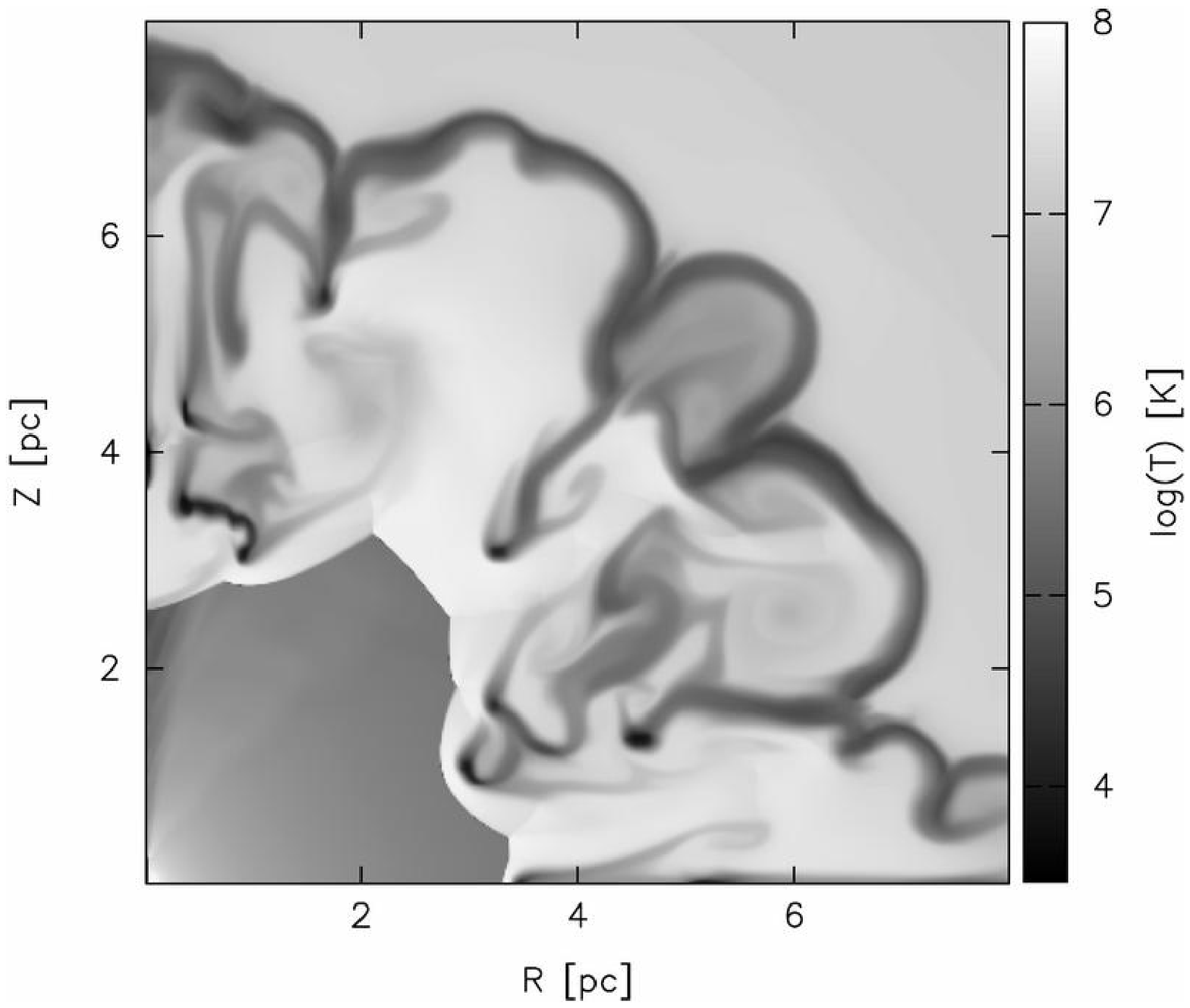}
\caption{Interaction of the Wolf-Rayet wind with the dense shell from
the red supergiant stage. Rayleigh-Taylor instabilities form as the WR
wind accelerates into the $r^{-2}$ density distribution left by the RSG
wind. The instabilities become enhanced when the WR wind collides with
the dense RSG shell and ruptures it. The panels show the temperature
of the gas after times (a) 15,000~years,
(b) 18,000~years (c) 21,000~years and (d) 24,000~years after the start of the WR wind.}
\label{fig:rt}
\end{figure}

Figure~\ref{fig:rt} shows the evolutionary sequence of the interaction
of the WR shell with the RSG shell. The dense, cold clumps formed due
to Rayleigh-Taylor instabilities in the WR shell act as obstacles
around which the hot, shocked WR wind has to flow. Large
amplitude ripples form in the RSG shell, through which the WR wind
eventually breaks out into the low-density bubble formed by the
main-sequence wind. The dense, cold clumps formed by the Rayleigh-Taylor
instabilities move outwards more slowly. These clumps will be subject
to hydrodynamic and photoablation but the numerical resolution of the
two-dimensional simulation is not able to follow these
processes. The mixing process produces gas with temperatures
$\sim 10^6$~K, which should emit X-rays. 

During the red supergiant phase, 16.5~$M_\odot$ of material
are lost from the star. All of this material is swept into clumps and
filaments by the Wolf-Rayet wind, which are potential
\textit{mass-loading} sources of the WR wind, to be discussed later.

\begin{figure}
\centering
\includegraphics[width=\textwidth]{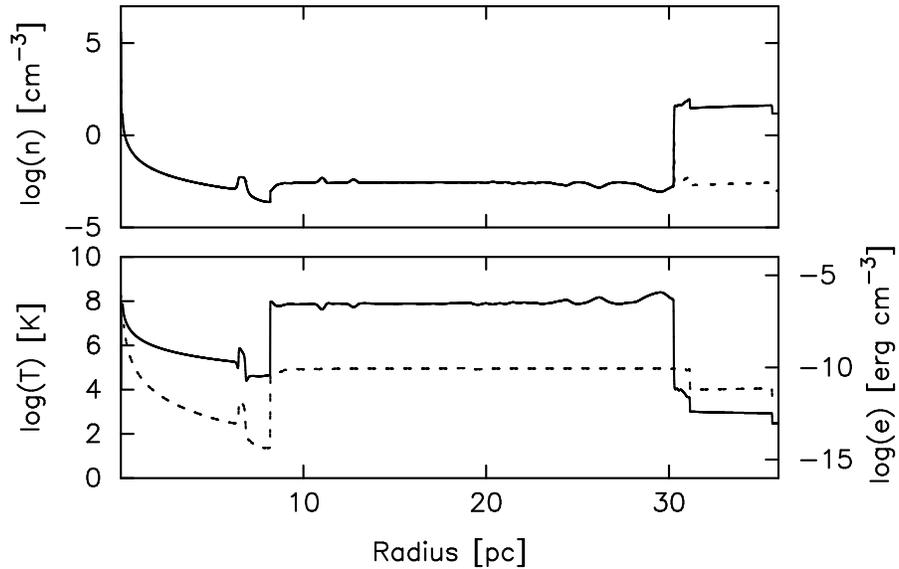}
\caption{Radial structure of a stellar wind bubble around a
40~$M_\odot$ star at the end of the Wolf-Rayet phase, i.e.\ after a
total time of
$4.8\times 10^6$~years. Top panel: total
number density (solid line) and ionized number density (dotted
line). Bottom panel: temperature (solid line) and internal energy ($e
= p/[\gamma - 1]$ --- dotted line).}
\label{fig:end}
\end{figure}

Figure~\ref{fig:end} shows the final radial structure of the wind
bubble around the $40 M_\odot$ star just before it explodes as a
supernova, from a one-dimensional (spherical symmetry) numerical
simulation. The low-density bubble has been repressurized by the hot,
shocked WR wind material, a new shock is being driven into the neutral
shell, and the H~II region is starting to reform. The one-dimensional simulation
cannot show the fate of the clumps and filaments formed during the
initial interaction between the WR wind and the RSG shell.

\subsection{Energy evolution in a wind-blown bubble}
\begin{figure}
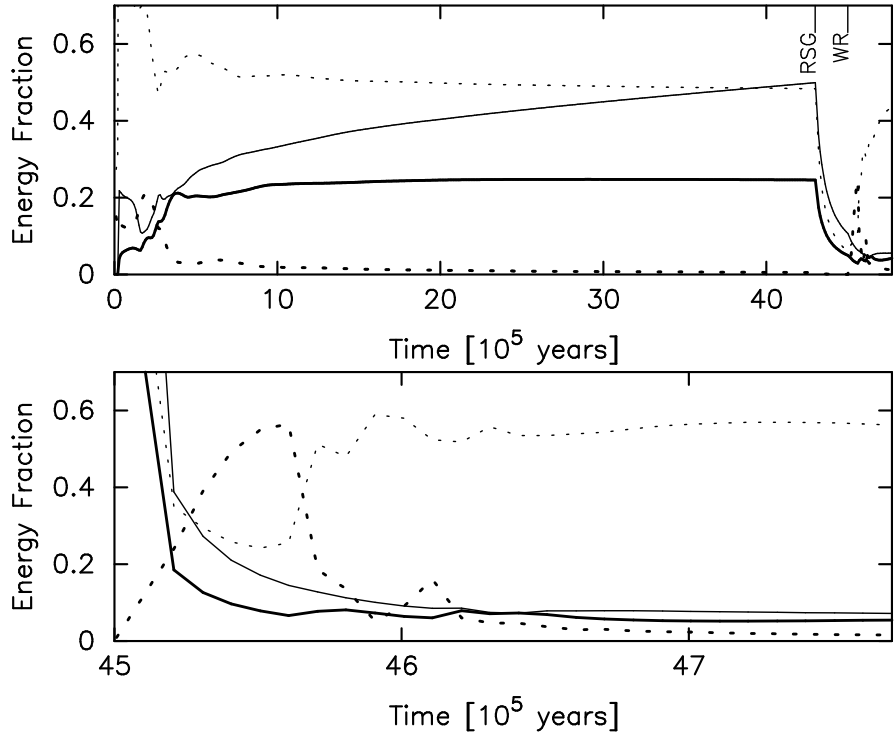

\centering
\includegraphics[width=\textwidth]{f9a}
\includegraphics[width=\textwidth]{f9b}
\caption{Top panel: Kinetic and thermal energies of the ionized and neutral gas
as fractions of the total stellar wind mechanical energy as a function
of time for the numerical simulation described in this paper. Thick
lines represent kinetic energy and thin lines represent thermal
energy, while solid lines represent neutral gas and dotted lines
represent ionized gas energy fractions. The onset of the RSG and WR
are indicated. Bottom panel: Kinetic and thermal energies as fractions
of the WR stellar wind mechanical energy, $\dot{E}_w (t - t_{\rm WR})$
during the WR phase.}
\label{fig:totenergy}
\end{figure}

Figure~\ref{fig:totenergy} shows the evolution of the neutral and
ionized gas kinetic and thermal energy fractions with respect to the
total stellar wind mechanical energy, $\dot{E}_w t$, resulting from
the numerical simulations described above.  During the main sequence
stage, the initial 400,000~yrs show large variations in the relative
importance of the kinetic energies in the neutral and ionized
gas. This is because this period of time sees the formation and
expansion of the initial H~II region, which then retreats back into the
shell of material swept up by the stellar wind before expanding
outwards again driven by the pressure in the hot bubble. The majority
of the main-sequence stage sees an energy-driven bubble in which the
thermal energy of the hot bubble pushes the expansion of a neutral
shell of material. The kinetic energy of the swept-up neutral shell is
roughly 0.25 of the stellar wind mechanical energy for most of this
stage. This is larger than the fraction $\frac{15}{77}$ predicted by
the simple thin-shell model \citep[see e.g.,][]{1997pism.book.....D}
and can be attributed to the H~II region. It is the H~II region that
results in the different
fractions summing to more than unity, since the contribution to the thermal
energy by the heating of the H~II region in the shell surrounding the
low-density bubble is substantial \citep[see also][]{2003ApJ...594..888F}.
Observations of only the ionized gas during this phase would show
$\epsilon \ll 0.1$, suggesting a momentum-driven flow, while
observations of the neutral shell would give $\epsilon > 0.1$, hence
an energy-driven flow.

When the red supergiant phase begins, the ionizing photons are trapped
close to the star and the outer H~II region recombines. Also, the
driving pressure in the hot bubble drops because the RSG stellar wind
is such low velocity. As a result, both the thermal and kinetic energy
fractions drop. Observations of both neutral and ionized gas during
this phase would give $\epsilon < 0.1$, suggesting a momentum-driven
flow.

During the Wolf-Rayet phase, the RSG wind material is ionized by the
high flux of ionizing photons and the high velocity WR wind forms a
hot, high-pressure bubble. This is reflected in the energy fractions
of the ionized gas, which rise steeply during this phase. It is a long
time before the wind-blown bubble becomes pressurized out to the
neutral shell, and so the neutral gas thermal and kinetic energy
fractions remain low.  Observations of the ionized gas during this
stage would show $\epsilon > 0.1$ initially, but later in the evolution
$\epsilon \ll 0.1$ even though the flow is now energy driven. The
neutral gas kinetic energy fraction corresponds to the outer swept-up
shell, which is not repressurized or accelerated during the timescale
we are considering. Consequently, observations of the neutral shell
during this stage of evolution would indicate that $\epsilon < 0.1$.

Figure~\ref{fig:totenergy} shows that neither observations of only the
ionized gas or only the neutral gas can reveal the full picture of the
energetics of a wind-blown bubble. Observations of the outer neutral
shell during this stage do not provide a true picture of what is
happening with the interaction between the current stellar wind and
its surroundings.

\subsection{Mass loading}
The presence of dense clumps embedded in a fast flowing stellar wind
and subject to a large flux of ionizing photons from the central
star suggests the possibility that material from the clumps will be
incorporated into the flow by one or other mass-loading processes,
namely photoevaporation, conductive evaporation or hydrodynamic
ablation. If the amount of mass incorporated into the flow becomes
important then the dynamics of the flow can be radically
altered. Both supersonic and subsonic flows tend to an average
transonic flow, and the positions of global shock waves
change \citep{1996ApJ...473..773S}. Furthermore, if the mass loading is strong enough, the density
in the hot shocked stellar wind can increase sufficiently that cooling
becomes important in this region.

Mass loading by embedded clumps in a stellar
wind was first proposed 
to explain the ionization potential-velocity correlation
observed in the WR ring nebula RCW~58 \citep{{1984MNRAS.211..679S}, {1986MNRAS.221..715H}}. In this scenario, the
mass-loading is due to hydrodynamic ablation and has two
regimes: one in which the flow around the clump is subsonic and the
other when it is supersonic. In the subsonic case, the mass-loading rate
is proportional to $M^{4/3}$, where $M$ is the flow Mach
number. In the supersonic case, the mass-loading rate saturates and is
taken to be constant. Numerical modelling of the mass-loaded flow in
RCW~58 shows that the position of the stellar wind shock is fixed by
the mass loading and that cooling occurs in the outer parts of the hot shocked bubble
\citep{{1993MNRAS.261..425A}, {1996A&A...313..897A}}. 

\section{Observations of wind-blown bubbles}
Bubbles around massive stars can be detected in a variety of ways: 
as optical ring-shaped nebulae \citep[e.g.,][]{1981ApJ...249..195C}, as shell-shaped
thermal radio continuum sources \cite[e.g.,][]{1995ApJ...439..637G}, as
neutral gas voids and expanding shells in the hydrogen 21~cm
line-emission distribution \citep{2005A&A...436..155C}, as infrared shells
\citep[e.g.,][]{1996AJ....112.2828M}, and in a limited number of cases as
diffuse X-ray sources \citep[e.g.,][]{1999A&A...343..599W}. Gamma-ray
burst afterglows probe the innermost regions of the progenitor wind
bubble \citep{2004ApJ...606..369C}, and Type II and Type Ib/c
supernova narrow spectral absorption features can be attributed to moving
shells around the massive star progenitor \citep{1984ApJ...287L..69D}.

Multiple concentric shells of material have been observed around many
galactic WR stars \citep{1996AJ....112.2828M}.\footnote{Ring nebulae have also been observed around
WR stars in the Large Magellanic Cloud \citep{1994ApJS...93..455D} but
their distance makes them less easy to study.}
 The outermost shells are thought to
correspond to the O star phase, and are most readily observable at far
infrared or 21~cm radio wavelengths, although around 8\% are seen
optically. These outer shells have can very large diameters ($>
100$~pc), depending on the ambient ISM density, and are expanding
slowly \citep[generally, $v_{\rm exp} < 10$~km~s$^{-1}$][]{1996AJ....112.2828M}. The innermost shells
are easily seen in optical narrowband images, and represent dense
material recently ejected from the star in a LBV or RSG phase or swept
up by the stellar wind in the WR phase.

Of the $\sim 150$ observed Galactic WR stars about 25\% are associated
with optical ring-like nebulae \citep{{1982ApJ...252..230H},
{1983ApJS...53..937C}, {1993ApJS...85..137M}, 
{1994ApJS...93..229M}, {1994ApJS...95..151M}}.  However, only $\sim 10$ of these WR ring
nebulae have the sharp rims and short dynamical ages (ring radius
divided by expansion velocity) that suggest that they are bubbles
formed by the winds of the central stars during the WR
phase.\footnote{These objects are: S~308, RCW~58, RCW~104, NGC~2359,
NGC~3199, NGC~6888, Anon (MR26), G2.4+1.4, and there is kinematical
evidence that the nebulae around WR~116 and WR~133 have expansion velocities consistent
with wind-blown bubbles \citep{1995A&A...304..491E}. Anon(WR~128) and
Anon(WR~134) are possibly wind-blown bubbles, too \citep{2000AJ....120.2670G}. There are also
wind-blown bubbles around two other evolved stars: NGC~6164-5 (around
an O6.5fp star), NGC~7635 (around an O6.5III star).}  The remaining
ring nebulae are less well defined and have dynamical ages much larger
than the lifetime of a WR phase, and hence are thought to be simply
stellar ejecta from the RSG or LBV stage photoionized by the WR star
(like a planetary nebula), rather than dynamically shaped by its
wind. Wind-blown bubbles were established as a class of object by
\citet{1981ApJ...249..195C} and \citet{1982Ap&SS..87..313L}.

\begin{figure}
\centering
\includegraphics[width=1.0\textwidth]{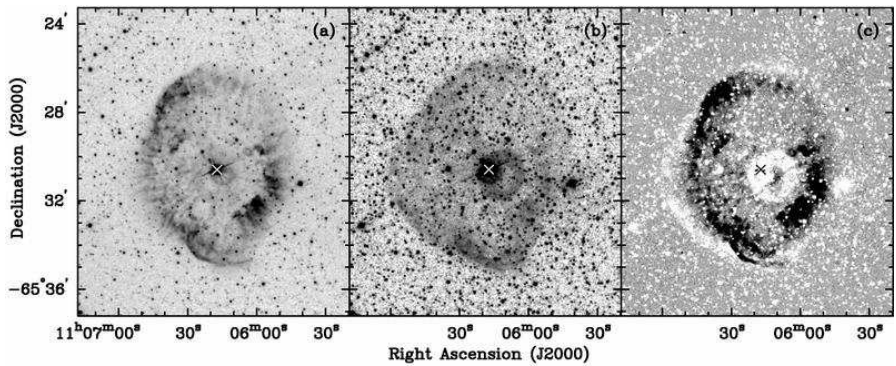}
\caption{(a) Optical H$\alpha$  and (b) [O~III] $\lambda 5007$ images of the
ring nebula RCW~58, together with (c) an image of their difference \citep{2000AJ....120.2670G}.} 
\label{fig:rcw58}
\end{figure}
A striking example of an optical WR ring nebula is RCW~58 (see
Figure~\ref{fig:rcw58}). This nebula consists of clumps and filaments
extending radially outward from the central WN8 star, seen in
H$\alpha$, with more diffuse [O~III] emission extending beyond
\citep{{1982ApJ...254..578C}, {2000AJ....120.2670G}}. The nebula is enriched in He and N, which
indicates that the material has been ejected from the star, probably
during a red supergiant stage. The expansion velocity of the shell is
87~km~s$^{-1}$, and broad linewidths indicate supersonic motions
\citep{1988MNRAS.234..625S}. The measured electron densities of the
clumps are $\sim$500~cm$^{-3}$.

Most WR wind-blown bubbles are aspherical to a lesser (e.g., RCW~58,
S~308, NGC~6888) or greater (e.g., NGC~2359, NGC~3199, G2.4+1.4)
degree. The asphericity is often more pronounced in H$\alpha$ images
than in [O~III], suggesting that it arises in the RSG or LBV ejecta
phase. Such morphologies could also be due to the environment: for
example, NGC~2359 and NGC~3199 are each bounded on one side
by molecular clouds
\citep{{1981ApJ...243..184S},{2001ApJ...563..875M}} and their roughly
parabolic shapes in H$\alpha$ images
could be the result of the wind-bubble blowing out from
near the surface of the dense cloud and expanding down a strong density
gradient, as has been suggested in the case of  G2.4+1.4
\citep{1990ApJ...359..419D}. An alternative explanation for this sort
of morphology could be that the star is moving supersonically through
the ambient medium and forms a bowshock ahead of it \citep[this was
suggested as an explanation for G2.4+1.4
by][]{1995MNRAS.273..443B}. 

\begin{figure}
\centering
\includegraphics[width=1.0\textwidth]{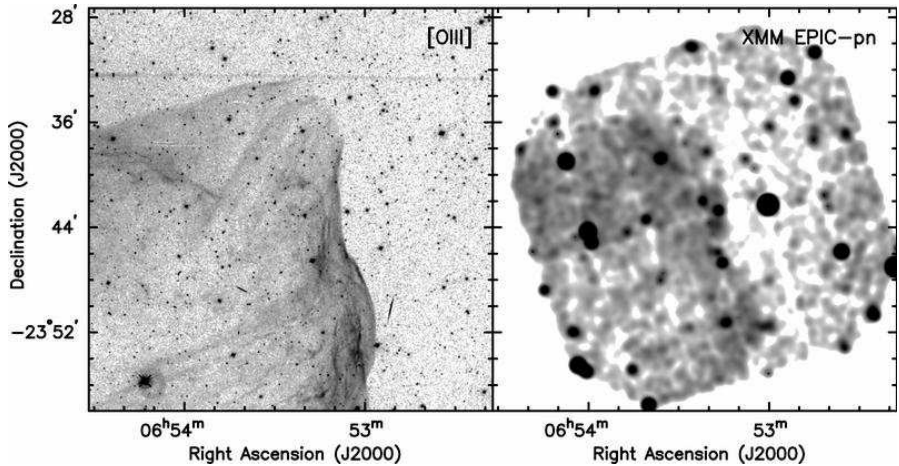}
\caption{Optical [O~III] $\lambda 5007$ image (left) and
\textit{XMM-Newton} EPIC image (right) of the northwest quadrant of
S~308 (\cite{2003RMxAC..15...62C}; figure reproduced with permission from
\textit{Revista Mexicana de Astrononom\'{\i}a y Astrof\'{\i}sica}).} 
\label{fig:chus308}
\end{figure}
\begin{figure}
\centering
\includegraphics[width=1.0\textwidth]{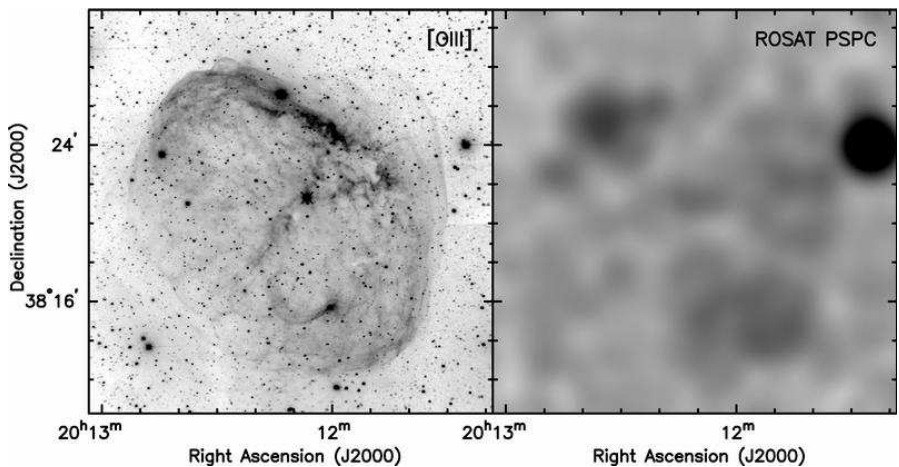}
\caption{Optical [O~III] $\lambda 5007$ image (left) and \textit{ROSAT} PSPC image (right)
of 
NGC~6888 \citep[][ figure reproduced with permission from
\textit{Revista Mexicana de Astronom\'{\i}a y Astrof\'{\i}sica}]{2003RMxAC..15...62C}.} 
\label{fig:chungc6888}
\end{figure}

Four of the 8 WR wind-blown bubbles have been observed
with X-ray satellites but only two of these, NGC~6888 and S~308, have
been detected in diffuse X-rays. Both were detected with the \textit{ROSAT} PSPC in the
energy range 0.1--2.4~keV and spectral analysis shows that the hot gas
in these bubbles is dominated by the component at $\sim 1.5 \times
10^6$~K \citep{{1988Natur.332..518B}, {1994A&A...286..219W}, {1999A&A...343..599W}}, although this could be
partially contaminated by higher energy point source emission. \textit{ASCA} SIS
observations of NGC~6888 suggests that there is an additional component of
hot gas at a temperature of $8 \times 10^6$~K \citep{1998LNP...506..425W,2005ApJ...633..248W}, while
\textit{XMM-Newton} EPIC observations of S~308 indicate that the
spectrum is very soft, suggesting plasma temperatures of only $1
\times 10^6$~K \citep{2003ApJ...599.1189C}. Figures~\ref{fig:chus308} and
\ref{fig:chungc6888} show the optical [O~III] $\lambda 5007$ and X-ray
images of S~308 and NGC~6888, taken from
\citet{2003RMxAC..15...62C}. The S~308 images clearly show that the
X-ray emission is limb-brightened, and NGC~6888 appears to be
limb-brightened also.\footnote{Although not clear from the \textit{ROSAT} image, this is evident in recent
\textit{Chandra} images:
\tt{http://www.chandra.harvard.edu/photo/2003/ngc6888}.}  In both cases the X-ray emission is interior to
the [O~III] emission, which marks the position of the main shock wave.

\section{Observations confront theory}
 It has been proposed that nebulae around massive stars begin with
 wind-blown shells in the main-sequence stage, evolve into amorphous
 H~II regions, then ringlike H~II regions, then nebulae with stellar
 ejecta in the RSG or LBV phase before the star explodes as a
 supernova
\citep{1984ApJ...278L.115M}. The bubbles blown by WR winds occur
within the circumstellar material of the RSG/LBV phase.  In practise,
wind-blown bubbles have rarely been observed around main-sequence O
stars
\citep[although kinematic evidence suggests that wind-blown bubbles do exist
inside H~II regions][]{2001AJ....122..921N}. Ringlike H~II regions
may be too faint to observe but expanding HI shells with masses up to
$10^4 M_\odot$ should occur around massive stars, as illustrated in
Figure~\ref{fig:ms}. In the later stages of
massive star evolution and, indeed, around some supernova remnants, 
expanding HI shells are detected \citep{2005A&A...436..155C} that must
have originated in the main-sequence stage. 

The fact that so few of the known WR stars are surrounded by
wind-blown bubbles suggests that this is a short-lived phase of
evolution \citep{1981ApJ...249..195C}. This is borne out by numerical
simulations, which predict observable bubbles only during the early
interaction of the WR wind with the LBV or RSG nebula
\citep{{2003ApJ...594..888F}, {2005astro.ph.12110F}}.  More common are amorphous
nebulae, which are observed around many WR stars, indicating a
photoionization rather than wind-blown origin, even when there are
clear indications that the central star possesses a strong stellar
wind. These cases could be the result of the environment in which the
nebula is found: if a strong density gradient is present the stellar
wind will blow out in the direction of decreasing density rather than
form a pressure-driven expanding bubble \citep[][discuss similar
models in the context of cometary H~II regions]{2005astro.ph.11035A}.
Clumps and filaments observed in nebulae such as RCW~58 are formed
naturally in numerical simulations by Rayleigh-Taylor and other
instabilities \citep{{1996A&A...305..229G},{1996A&A...316..133G}}.

Theoretical estimates of X-ray emission as predicted by analytic
theory fail to match the reality. Models that do not include thermal
conduction have very high temperatures in the shocked stellar wind and
do not produce the soft X-ray spectrum that has been
detected in NGC~6888 and S~308. On the other hand, the self-similar
models of \citet{1977ApJ...218..377W}  and
\citet{1995ApJ...455..145G}, which include thermal conduction, 
predict much higher values of the soft X-ray luminosities than are
observed. It appears that some sort of dissipative process, though not
the classical thermal conduction proposed by
\citet{1977ApJ...218..377W}, is necessary to produce the soft
X-rays. This argument presupposes that the X-ray emission is produced
in the shocked Wolf-Rayet wind material. Two-dimensional numerical
simulations of the wind-blown bubble produced by a $35 M_\odot$ star
\citet{2005astro.ph.12110F}, on the other hand, show that the X-ray
emission actually arises in the shocked, dense RSG wind material. This
can be appreciated in Figure~\ref{fig:rt}, where gas temperatures of
$\sim 10^6$~K, characteristic of soft X-rays are present in the
shocked RSG shell material. Also, it could be expected that material
photoevaporated or ablated from the dense clumps formed by the
instabilities could augment the soft X-ray luminosity in the
interaction region. 

\section{Final Remarks}
Although the general characteristics of the formation and evolution of
wind-blown bubbles around evolved stars are well understood, there
remain a number of important features to explain. Most obviously, it
is interesting to know whether the asphericity of nebulae such as
NGC~6888 and RCW~58 is due to the non-spherical ejection of material
in the RSG or LBV stage, or whether it is due to some
direction-dependent physical process, such as thermal conduction, in
the WR bubble phase. Photoionization is clearly important for the
thermal balance and energetics of wind-blown bubbles and should not be
neglected in numerical calculations.  The role of instabilities in producing the clumps
and filaments observed in some (but not all) bubbles should be
examined in more detail. Cooling within the clumps can lead to the
formation of very
dense, neutral structures within the hot, ionized bubble. The
ablation, either by photoevaporation, thermal conduction or
hydrodynamic ablation, of this neutral material will modify the
physical properties and chemical abundances of the hot, shocked wind and such processes should,
therefore, be studied further. Theories should be tested with
real data and new observations will always be needed, at all
wavelengths, to provide more detailed kinematical information,
abundance data, and improved stellar wind parameters.

\end{document}